# Analyzing ENUM Service and Administration from the Bottom Up: The addressing system for IP telephony and beyond.

Junseok Hwang, Milton Mueller, Gunyoung Yoon, and Joonmin Kim
jshwang, mueller, gyoon, jmkim@syr.edu
Graduate Telecommunications and Network Management Program
School of Information Studies
Syracuse University, Syracuse, NY 13244-4100
Tel) 315-443-4473, Fax) 315-443-5806

September 21, 2001

**Abstract**

ENUM creates many new market opportunities and raises several important policy issues related to the implementation and administration of the ENUM database and services. ITU's recent World Telecommunications Policy Forum 2001 dealt with the emergence of ENUM as an important numbering issue of IP telephony. This paper prepares some important emerging issues of ENUM administration and policy by taking an empirical research approach from the bottom up.

We will identify potential key ENUM services, and estimating the size of the service market opportunities created by the availability of PSTN-IP addressing and mapping mechanisms, particularly in the context of IP telephony. Also, we analyze the possible administrative models and relationship scenarios among different ENUM players such as Registry(ies), Registrars, Telephone Service Providers, ENUM Application Service Providers, etc. Then, we will assess the effects of various administrative model architectures of ENUM service by looking at the market opportunities and motivations of the players. From the empirical findings, we will draw the implications on transactions among different kinds of ENUM service providers. Finally, the results of the model analysis will be used for the discussion of policy related issues around the ENUM and IP telephony services.

## 1 Introduction

Should we be numbering services or users instead of phone lines?

Recently, the IETF standardized the ENUM protocol [Faltsrom, 2000]. The ENUM protocol uses Internet Domain Name System (DNS) based naming and resolution mechanisms to map E.164 telephony numbers into other numbering addresses, especially IP based addresses (URLs, E-mail address, IP phone, Call setup signaling server, etc.). Accordingly, ENUM creates addressing mechanisms for IP telephony and also enables new services using E.164 numbers as integrated identifiers for users and/or services. Multi-service gateway applications will be available for both telephony and the Internet using addressing technology like ENUM.

The main advantage of ENUM is the converged IP-PSTN network that will exploit the network externality and utilizing one phone number for any other identifiers. For IP users, it connects IP-based terminals to cheap and ubiquitous PSTN phones, while bypassing costly PSTN long distance circuits. It also allows PSTN users to interwork with IP-based networks by allowing PSTN terminals to reach both PSTN and IP network users.

Therefore, ENUM will create many new market opportunities. Already various market players are avidly pursuing them. Also, it raises important policy issues related to the implementation and administration of the ENUM database and services. As such it is receiving significant attention from telecommunication regulators and policy makers at the national and international level. ITU's recent World Telecommunications Policy Forum 2001 dealt with the emergence of ENUM as an important numbering issue of IP telephony [WTPF, 2001]. Policy issues frequently discussed include the impact of ENUM on carrier selection, bypass, interconnection, registry competition, registrar eligibility and numbering authority for ENUM identifier. This paper analyzes these important emerging issues of ENUM provisioning, administration and policy by taking an empirical research approach.

Most discussions of ENUM proceed from the top down. They discuss the architecture and technical capabilities of the service and the organization of authority over the database. This paper uses a bottom-up approach to analyze the issues. We begin by identifying potential ENUM services and estimating the size of the associated service market opportunities, focusing particularly on IP telephony. Identifying the additional market potential available to new ENUM-based services for different players in the related industry, we analyze the possible administrative models and relationship scenarios among different ENUM players such as Registry(ies), Registrars, Telephone Service Providers, ENUM Application Service Providers, etc.. We generalize the recently proposed IETF ENUM administrative models [Pfautz, 2001] to represent various ENUM-based services and administration scenarios. Then, we assess the effects of various administrative model architectures of ENUM service by looking at the market opportunities and motivations. From the empirical findings, we will draw the implications on transactions among different kinds of ENUM service providers. Finally, the results of the model analysis will be used for the discussion of policy related issues around the ENUM and IP telephony services.

## 2 Brief Background and Overview

The purpose of this section is to outline the key functionalities and operations of ENUM. More definitive discussions can be found elsewhere.

## 2.1 ENUM Basic Functionality

ENUM is a name devised by the Internet Engineering Task Force (IETF) to refer to the protocol (RFC 2916, "E164 number and DNS") designed by its Telephone Number Mapping working group. Based on Domain Name System (DNS) operation, ENUM protocol provides the functionalities of the resolution of a telephone number to a Uniform Resource Identifier (URI) that can relate a range of service resources. ENUM can provide services connecting E.164 telephone numbers to Internet telephony-oriented services by enabling telephone users or entities to access Internet based resources. It makes extensive use of Naming Authority Pointer Records (NAPTR) defined in RFC2915 to identify available ways or services for contacting a specific party identified by the E.164 number.

## 2.2 ENUM E.164 Number Resolution Scheme: RFC2916 [Faltsrom, 2000]

In RFC 2916, the domain "e164.arpa" is proposed as the root of the DNS naming hierarchy for storage of E.164 numbers. This domain is further divided into sub domains to facilitate distributed operations. E.164 number resolution is the process of finding the DNS name for a specific E.164 number. It is briefly described as follows:

> 1. Start with a complete E.164 telephone number.
> Example: +1-315-443-4473
>
> 2. Remove all characters other than digits except for the leading "+", which is to flag that the number in query is an E.164 number.
> Example: +13154434473
>
> 3. Remove all characters other than digits.
> Example: 13154434473
>
> 4. Insert a "." between each digit and at the end.
> Example: 1.3.1.5.4.4.3.4.4.7.3
>
> 5. Reverse the order of all the digits.
> Example: 3.7.4.4.3.4.4.5.1.3.1
>
> 6. Append, as a suffix, the string of "e.164.arpa".
> Example: 3.7.4.4.3.4.4.5.1.3.1.e164.arpa.

There can be two potential implementations of ENUM domain names. RFC 2916 suggests that the domain names standardized under e164.arpa, which will result in one ENUM root structure. However, multiple domain implementation rooted under different domain names would be possible in the market place if there is no regulatory restriction or policy regarding ENUM domain policy. We just named those domains as "any" in the illustration of Figure 1. The competitive multiple ENUM root implementation would require associated ENUM registry discovery mechanisms and peering interconnection of registries and .arpa domain implementation will be managed on the planned and provisioned hierarchy of the domain resources. [ITAC-T SGA ENUM, 2001]

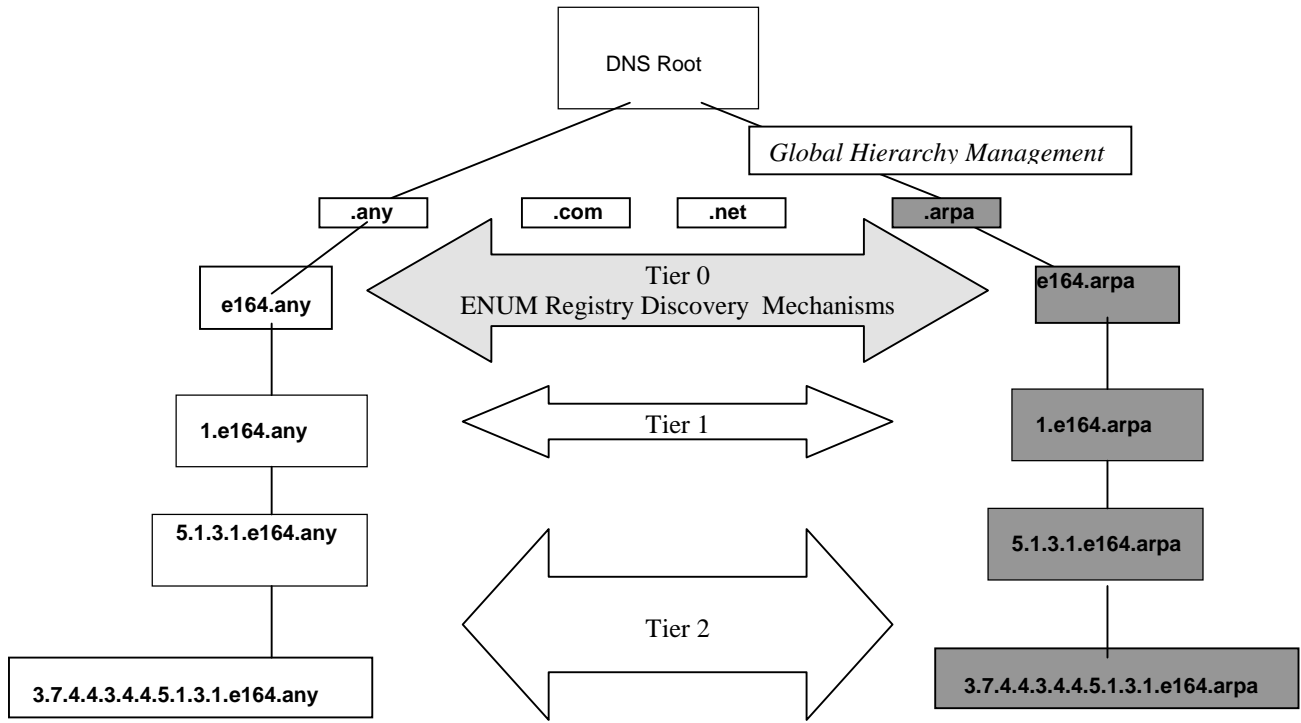

Figure 1: ENUM Domain Potentials

**2.3 Market Players for ENUM Number Resolution**

There are a number of different ENUM provisioning market models proposed by different organizations. Most of them adopt similar structures that contain registry, registrar, and ENUM Application Service Providers (ASPs), with different views of who and how they will participate in each entity. Here we suggest generic version of ENUM market players and their functions.

- **Pointer to ENUM Registry (Tier-0)**
This will point to a Registry that is an authoritative name server for that country code or portion of a country code. Internet Architecture Board recommend the Reseaux IP European Network Control Center (RIPE-NCC) maintains the e164.arpa implementation. However, for the competitive and multiple registry roots, there is no such centralized database suggested yet.

- **National Registry (Tier-1)**

  Registry (ies) will manage name servers whose entries point to Service Registrar for a Telephone number. Candidates include domain name registry, telephone number registry, or a new provisioning body. There are two competing models:
  - A regulated national monopoly administrative structure
  - An open level competitive administrative structure

- **Registrar (Tier-2)**

  Registrars database providers that host NAPTR Records for a Telephone number. Competing model players include:
  - Any properly accredited domain name registrar
  - TNAAs (Telephone Number Assignment Authorities)
  - Telephone Service Provider (TSP)
  - ENUM Application Service Providers: Any organizations or enterprises that provide ENUM related services such as Internet Telephony Service Providers (ITSPs), ISPs, and other web-based service providers.

In the following, the players related to the ENUM provisioning model are categorized as the groups that share the same business motivations. Since telephone companies play several roles as E.164 Number Providers, traditional telephony service providers, and facilities-based service providers, the TSP's motivation in ENUM implementation and provisioning model will depend on each of these roles.

- **End Users**

  End users are the consumers that include not only the individual users but also businesses and organizations that need services either from the ASPs or directly from the Facilities-based Service Providers (FSPs).

- **Facilities-based Service Providers (FSPs)**

  Services include fixed telephony, private leased circuits, mobile services, Internet connections, and other services. TSPs (ILECs amd CLECs) and ISPs are included in a sense that they own the facilities and the network. ISPs are categorized as Facility-based Service Providers whether they own and operate the network or they lease. ISPs are considered to be separate from the Application Service Providers (ASPs).

- **Application Service Providers (ASPs)**

  ASPs include ITSPs and other potential ENUM application service providers. Services include voice resale, IP Telephony, International callback, aggregator, rebiller, in-building communication services and other value-added telecommunication services. Voice Portals, that provide applications such as call center/web integration and voice powered e-commerce, are also included as ASPs. They will become an important source for mobile users. New voice services will complement microbrowsers, helping get data into and out of small handheld devices.

# 3 Key ENUM Service Market Potential: Bottom-Up

## 3.1 Major Potential Services

### 3.1.1 Basic IP telephony service potential

IP Telephony is likely to be the first application of ENUM that will drive the market. Since it is the basic service that will take a full advantage for general users who are connected in both PSTN and IP network, current ENUM issues such as administration and policy analysis can be based on this service. Currently there are many IP Telephony services available, but with critical limitations such as limited user base for each service provider and no service or different dialing for phone-to-PC connection. Once the provisioning model is developed, ENUM will solve these problems without significant technical challenge by maintaining a single logical database for all the telephone numbers and mapping them to other URIs.

### 3.1.2 Value added services potential with added functions

Many of the services that use simple text format may be interoperable by text-voice converting services. For example, a telephone user can send email to someone on an IP network using a voice ↔ text service. It may also be possible to develop some new terminal devices that connect via the PSTN to integrate the two networks, such as PDAs with cellphone functions, IP video conferencing, and Web TV. However, revolutionary changes in the type of access devices used would eliminate the advantage of ENUM, which is its ability to connect the installed base of telephone users to other networks. Unified Messaging Service (UMS) is an example in between the basic service and services with a moderate degree of added functions. In addition to existing services, new services for specific user group such as intranet in an organization may be developed. Call forwarding is an example. To develop a specific service that would be technically feasible and profitable at the same time, each service should be examined in detail. Moving onto the second generation of IP Telephony, Intelligent Network (IN) services such as 800 toll-free service, local number portability, and calling card services are complemented in IP networks as well as PSTN, by using Signaling System 7 (SS7)/ IP gateways supporting IP Telephony switches. As IP Telephony matures, those services will be expected as basic on top of the internetworking rather than creating a new market. However, consolidated NAPTR Resource Records of ENUM may enable new services that manage bundled accounts and billings that will be offered by ASPs as enhanced services.

## 3.2 Major Potential Service Market Estimation and Opportunities

### 3.2.2 IP Telephony

Low cost is not the only incentive for IP telephony with today's low rate for circuit-switched call. Both individual and business consumers want new applications, high bandwidth, business solutions, reliability and QoS (Quality of Service). Although they do not have strong buying power because of the high demand in data communication needs, as the competition moves from price to new services, they want more multimedia applications and Session Initiation Protocol (SIP) based services such as web-integrated call centers, multimedia conferencing, IP Conferencing, unified messaging etc. Multinational corporations can save 40% on international and domestic calls by integrating voice and data on a single IP network. Smaller companies also benefit from managing a single network and applications. Many forecasts show IP Telephony exploding in the near future, though one should be cautious of hype. IP LAN Telephony market will increase by almost 90 percent per year for the next five years, from $138M in 2000 to $3.2B by 2005. [PR Newswire, 2001] Since end-users want both new services and integration, they will have a strong incentive to subscribe to ENUM services as long as the incremental cost is not substantial.

|  |  | 2000 | 2001 | 2002 | 2003 | 2004 |
|---|---|---|---|---|---|---|
| Internet Users, (k) | World | 177,464 | 239,939 | 311,194 | 391,500 | 495,685 |
|  | % Growth | - | 35.2 | 29.7 | 25.8 | 26.6 |
|  | USA | 76,500 | 95,414 | 118,679 | 137,849 | 155,581 |
|  | % Growth | - | 24.7 | 24.4 | 16.2 | 12.9 |
| PC-to-Phone Users, (k) | World | 5,137 | 11,324 | 22,394 | 40,364 | 69,328 |
|  | % of Internet Users | 2.89 | 4.72 | 7.2 | 10.31 | 13.99 |
|  | USA | 2,861 | 5,710 | 10,624 | 17,427 | 26,439 |
|  | % of Internet Users | 3.74 | 5.98 | 8.95 | 12.64 | 16.99 |
| Usage, (k mins) | World | 3,361 | 10,920 | 29,614 | 69,036 | 139,634 |
|  | USA | 1,806 | 5,924 | 15,578 | 33,956 | 63,712 |
| Market Size, ($M) | World | 185 | 512 | 1,201 | 2,431 | 4,190 |
|  | USA | 87 | 234 | 517 | 948 | 1,478 |

Fig. 3-1 IP Telephony: Exploiting Market Opportunities [IPMARKET, 2001]

Fig. 3-1 shows forecasts of the growth of Internet users and PC-to-Phone users. The numbers imply that the users are growing at a rate of 25-30% annually. The potential market for ENUM, however, might grow even faster because it will enable phone-to-PC calls, and hence include all the telephone users and wireless users. Furthermore, it will also stimulate the existing PC-to-phone calls as well as the overall Internet uses, especially broadband connections. It is expected that after 2003, more than 10% of the Internet users worldwide will use the Internet to call regular PSTN phones. Wireless is another rapidly growing market and the 3rd generation wireless will take a full advantage of ENUM with its data communication-ready interface. Fig. 3-2 shows the potential ENUM market (IP telephony call to the Internet Phones and PCs) in revenue and the number of subscribers that includes phone lines and mobile phones and total Internet use. We assumed 5% penetration for such users as the potential market. The USA market estimates for 2002 are based on the growing rate of the world market and the portion of the USA market to the world market. With such assumptions, we could derive the potential market generated for phone-PC IP telephony will have about 25M subscribers and $11B worldwide market potential at the end of year 2002. Assuming Phone-to-PC and PC-phone users are same users, we can draw an estimation that the potential IP-

telephony user base for ENUM addressing number service could reach upto 25M subscribers by the end of the year 2002.

### 3.2.2 Unified Messaging Service (UMS)

Unified Messaging Service may be considered as an extended application to IP Telephony that will utilize ENUM service as well. However, it has a potential to provide much broader range of services if integrated with other technologies such as VXML, speech recognition, and text-to-voice. [Datacomm, 2000] Voice, email, fax, and other multimedia messages will be accessed and updated through one server directly to centralized universal directories in the future UM message process, instead of using different server for each type of messages and then transferring them into the directories.

### 3.3 ENUM Players in Major Potential Service Markets

In this section, we will define and describe the major corporate players at Tier 1 and Tier 2 who will shape the interface between ENUM and IP telephony service.

### 3.3.2 TSPs –Incumbents (ILECs)

While the economic advantage of IP Telephony drove the growing market, IP Telephony has been a threat to established voice carriers, especially to the long distance carriers (IXCs). Their existing revenue streams may be cannibalized by a shift to VoIP. Figure 3-4 shows that long distance telephone revenue was already declining in the past two years and is expected to continue to do so in the future. Even if VoIP is cheaper for carriers on the supply side, it may not be economically rational to move immediately towards providing telephony services over an IP platform. The speed of transition will depend on the degree of competition and whether a particular carrier is an incumbent or a new market entrant. [IPTelephonyWorkshop,2000]

Incumbent carriers that already have existing revenue streams and the traditional network will gradually deploy IP Telephony for the following benefits:

- More efficient use of bandwidth (64 kbps for traditional vs. 6 kbps for IP call by codec used to snip out the pauses and quiet parts of speech)

- Market for corporate VoIP (17% of US businesses began the implementation of IP LAN Telephony in 2000, will grow to 80% in four years [Philips])

- Operators in developing countries may be better advised to embrace IP Telephony, and bear the consequences of reduced per-minute revenues from long-distance and international services, than to risk missing the opportunity to develop revenues in future growth areas [IP TelephonyWorkshop, 2000].

- Lower revenues from international traffic may be compensated through higher revenues from Internet access services, from selling more network capacity to users such as ISDN and second line, and from higher volume of local Internet calls. It could also stimulate people to make more international calls, via Internet but also the PSTN. [OECD, 1998]

| US$ billions | *2000* | | *2002* | |
|---|---|---|---|---|
| | World# | USA | World# | USA |
| **Telecom Services market revenue (current prices and exchange rates)** | | | | |
| Total Toll (Estimates by 50% of Total Telephone Revenues) | 230 | 108 | 205 | 98 |
| Telephone (Total) | *460* | *220('99)\** | *410* | *~195* |
| Mobile | *230* | *48('99)\** | *315* | *~65* |
| Other | *150* | *45^* | *200* | *60^* |
| Total | *840* | *313* | *925* | *320* |
| **Other statistics** | | | | |
| Main telephone lines (millions) | *970* | *184('99)+* | *1'115* | *~211* |
| Mobile cellular subscribers (millions) | *650* | *86('99)+* | *1'000* | *~135* |
| Personal computers (millions) | *500* | *161+* | *670* | *~215* |
| Internet users (millions) | *311* | *95+* | *500* | *~152* |
| # ITU, Apr. 7, 2000 APPENDIX A<br>\* FCC Release, Dec. 21, 2000 APPENDIX B<br>'+ ITU, 2001 APPENDIX C<br>^ Estimates by % of Subscribers (USA/World=29.31%, FCC)<br>` FCC, 2000 APPENDIX D | | | | |
| **Phone-to-PC Market Potential Revenue ($B) – 5% penetration for potential market** | *30.5* | *10.05* | *36* | *11.15* |
| Local, Long Distance, and International Toll Revenues plus Mobile and Other Revenues | | | | |
| **Phone-to-PC Market Potential subscribers (M) – 5% penetration for potential market** | *96.55* | *17.75* | *130.75* | *24.9* |
| Total number of Internet users including broadband connections plus Total number of PSTN users including mobile cellular | | | | |

Fig. 3-2 Potential global market for ENUM addressing for IP telephony

| | | 2000 | 2001 | 2002 | 2003 | 2004 | 2005 |
|---|---|---|---|---|---|---|---|
| Number of Mailboxes (K) | World | 477 | 2195 | 5470 | 12416 | 22601 | 42812 |
| | North America | 346 | 1386 | 2868 | 6245 | 10338 | 19325 |
| Revenue ($M) | World | 1201 | 2810 | 5459 | 8434 | 12463 | 18210 |
| | North America | 766 | 1490 | 2762 | 3713 | 5152 | 7672 |

Fig. 3-3 UM Service Forecast, Ovum Forecast, 2000 [UMS, 2000]

Following figures show the cost comparison of IP call and switched call to TSPs and their spending on the IP telephony deployment respectively. The numbers indicate that, despite the various speed of deployment, TSPs are moving into the IP-based network for their services in the future. As the incumbent TSPs are moving toward the IP-based services, the importance and benefits of ENUM for number mapping and address resolution will increase. The ILECs will have incentives to create their own (probably) resolution architecture for their potential market due to the increase of IP-based services. Therefore, expected strategies from ILECs on ENUM resolution policy would be the position where they can keep their numbering resource resolution capability no matter which ENUM service provider the ENUM registrant choose.

### 3.3.3 TSPs –New Market Entrants (CLECs and IXCs)

New local telephony market entrants such as CLECs and IXCs will be more advantageous because they own the facilities for voice, data, and Internet access with modern networks; hence they can provide converged services with IP telephony and ENUM. They can also price more freely than a regulated monopoly, and they can bundle local and long-distance services ahead of the incumbent carriers. [Computer Reseller, 1998] Currently, most new entrants are focusing on specific client types such as commercial businesses instead of home subscribers. As such, CLECs are growing rapidly in metropolitan regions while not maintaining any network infrastructure between regions. With IP telephony, the both CLECs and IXCs can easily connect their regions together over a private IP network allowing low cost intra-network calling. IP telephony gateways can become the glue to merge their metro PSTN networks together instead of relying on other PSTN service provider who charge an excessive rate for access. Other advantages that the new market entrants have include lower cost for infrastructure (equipment cost less and occupies 90 percent less space) and faster market entrance. More of CLECs provide the IP-based services and they are evidently motivated to utilize ENUM enabled service infrastructure. Having ENUM based address management, CLECs and IXCs would be able to have more independence from ILECs of providing the basic and enhanced IP telephony services. This will motivate new entrant TSPs to develop a capability in the ENUM numbering resolution.

### 3.3.4 ISPs

Another possible player in ENUM or IP telephony architecture is the Internet service provider. However, if the Internet were to be connected to the telephone system in a widespread way, it is not clear who would install, operate and benefit from the gateways. As mentioned previously, CLECs and IXCs will benefit directly from the gateways. For ISPs, it requires a way to tie the lower-level Internet service to the higher-level IP Telephony service, an example of vertical integration in the Internet industry. [Clark, 1997] Several IP telephony cost analysis shows that transport cost is the largest part of total cost to deploy Internet Telephony for both ISPs and TSPs. [Leida, 1997][Weiss, 1998] The competitive advantage of an ISP would be efficiency of operating its network, such as economies of scale, facilities-based networks, and network optimization techniques. In any case, whoever installs, operates, and provides the gateway services must bill someone for the use. Anyone who will use and operate those gateways will have incentives utilizing the ENUM capability for their customers.

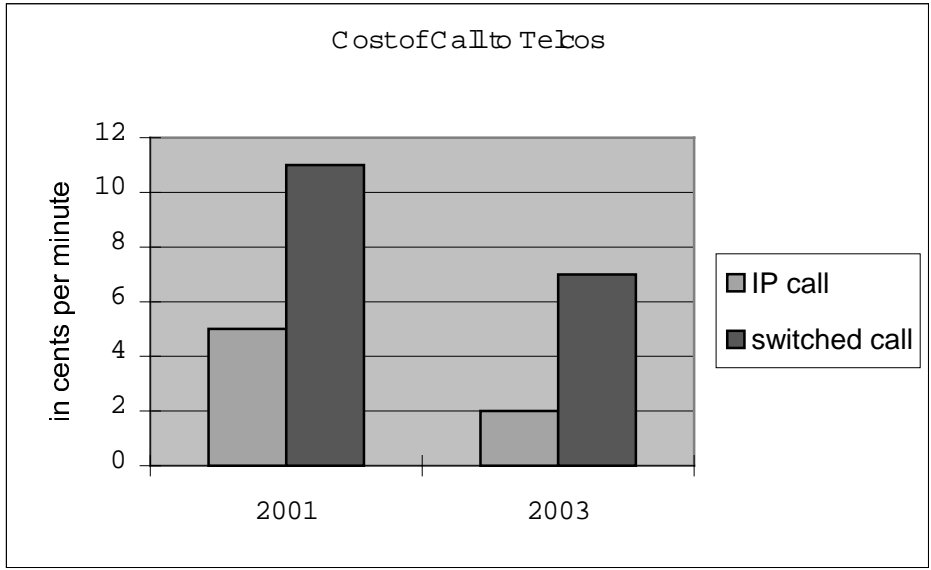

Figure 3-4 Cost of Call to Telcos [The economist, 2001]

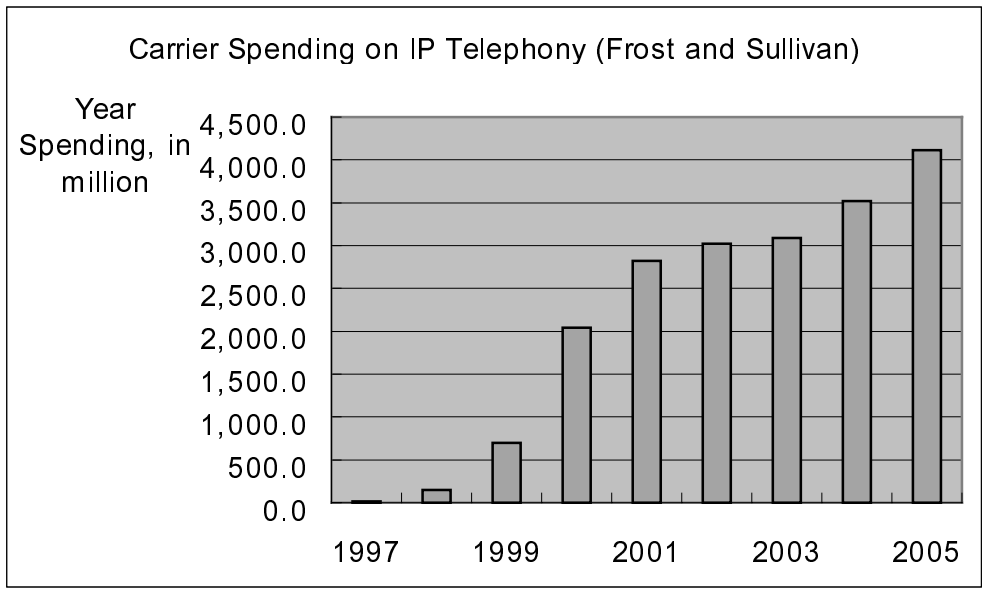

Figure 3-5 Carrier Spending on IP Telephony [Frost et al.]

### 3.3.5 ASPs

Finally, ENUM Application Service Providers (ASPs) may be categorized as those who does not own any facility based transport and switching network, but have revenues from various applications related to ENUM service capability. Potential ASPs including Internet Telephony Service Providers, Corporations, Call Centers, and Service Bureaus are the ones that directly generate the revenues from the service; hence their motivation is relatively clear. The success of each of them would be decided by the market through free competition, and greatly dependant on the profit model and how innovative the applications they offer. They will have incentives to reach the end customers with ENUM enabled services as Tier 2 of the ENUM services.

## 4 ENUM Administration Market Reference Model Analysis

In this section, we will set out six different administration models for ENUM and examine their relationship to the key players described in Section 3.3. Depending on the ENUM administration functions of the players, there could be several administration models and scenarios. For example, we can expect a monopolistic infrastructure or open competitive market for ENUM resolution service according to proposed administration models. Since the ENUM resolution process is combined with the domain name registration process and the carrier selection process, the relationships are complex. Further, the implementation and administration of ENUM are closely related to policy and regulation issues.

### 4.1 Model Assumptions

As this is a first order analysis, there are some simplifying assumptions that we wish to make. These include:

- The Tier 1 entity, i.e., the registry, handles the sub-domain of e.164.arpa corresponding to a particular country code and points each served telephone number to Tier 2. We did not make any restriction who can be the Tier 1 registry service providers. The registry knows the Telephone Number Assignment Authority and TSP for each telephone number. We assume that a registry discovery mechanism is in place when multiple registries and roots are implemented.

- In Tier 2, the entity that acts as a registrar could be a Telephone Service Provider (TSP), an Internet Service Provider (ISP), an Application Service Provider (ASP) or an independent registrar. Tier 2 registrars must control the access rights to the NAPTR RRs so that TSP, ISP, ASP or Independent registrar can provision, access and change their own NAPTR RRs. For example, a TSP has authority over provisioning of the NAPTR RRs at the Tier 2 for network-related services. All Naming Authority Pointer Resource Records (NAPTR RRs) for a given telephone number must be stored in a registrar's name server.

- Telephone Service Providers (TSP) may offer the same application service to users as Application Service Providers (ASPs).

- The ENUM subscriber can select any registrar to use for its assigned telephone number. The user can use the ENUM service as a personal number service and can subscribe to many application services from various providers, including his/her Telephone Service Providers.

- The ENUM subscriber can change the registrar without informing the current serving registrar. In this case, the new registrar may inform the old registrar and also inform registry about registrar change for the telephone number.

- The ENUM subscriber can lose his/her rights to the assigned telephone number and ENUM services when he/she disconnects the telephone service and no longer owns the telephone number. However, if he/she wants to have the telephone number only for using telephone service, he/ she can disconnect just ENUM service.

## 4.2 Six Reference Model Analyses

Industries for ENUM-based services can be developed effectively with the appropriate policy on the administration models for different ENUM players. The industry and policy bodies are recently working over various scenarios of tier 1 implementation and tier 2 operation to come up with the consensus on appropriate ENUM administration policy. Different players have different business interest and incentives to have the presence in the ENUM services. It is not an easy task to identify a model that satisfy all different players' interests. There was a previous study on the ENUM administration process and reference model by Pfautz et al. [Pfautz, 2001]. The study was based on the assumption of single registry scenario and focused on the technical interface of the ENUM administration process. In this paper, we examine the ENUM administration models for both single registry and multiple registry scenarios and further focus on the business industry structure and incentives of different administration scenario models.
The analyses of different models will be conducted to draw the implications to address important telecommunication policy issues surrounding ENUM services and markets.

Six reference models are presented here. Model 1,2,3 are based on one registry that can make authoritative one, and other models from 4 to 6 have multiple registries (both sub-layer and independent layer approach) of competitive one. According to the role of Tier 2 entities, functions among entities could be changed.

### 4.2.1   Model 1

In this first model, the Telephone Service Provider (TSP) that currently provides the telephone service to the telephone number is the registrar, and there could be other registrars that compete on equal terms. In addition to the principal role to assign telephone number to the use for telephony service, the TSP can act as the ASP for the users' application service by bundling the ENUM enabled applications into their existing services. However, the user can decide which ASP to use. The TSP already has a service relationship with the user so that it would be easier for the TSP to verify both the user's identity and the telephone number assignment to that the user to authenticate to provisioning request. The TSP will charge the end-users for ENUM registrar service and the TSP will pay the Registry for its registry database services. The registry service charge will be generally the flat fee for each number. The TSPs can provide ENUM service as part of their enhanced services. Also, TSPs might have chance to provide ENUM subscription service with the assigned numbers for those who do not subscriber actual telephone service.

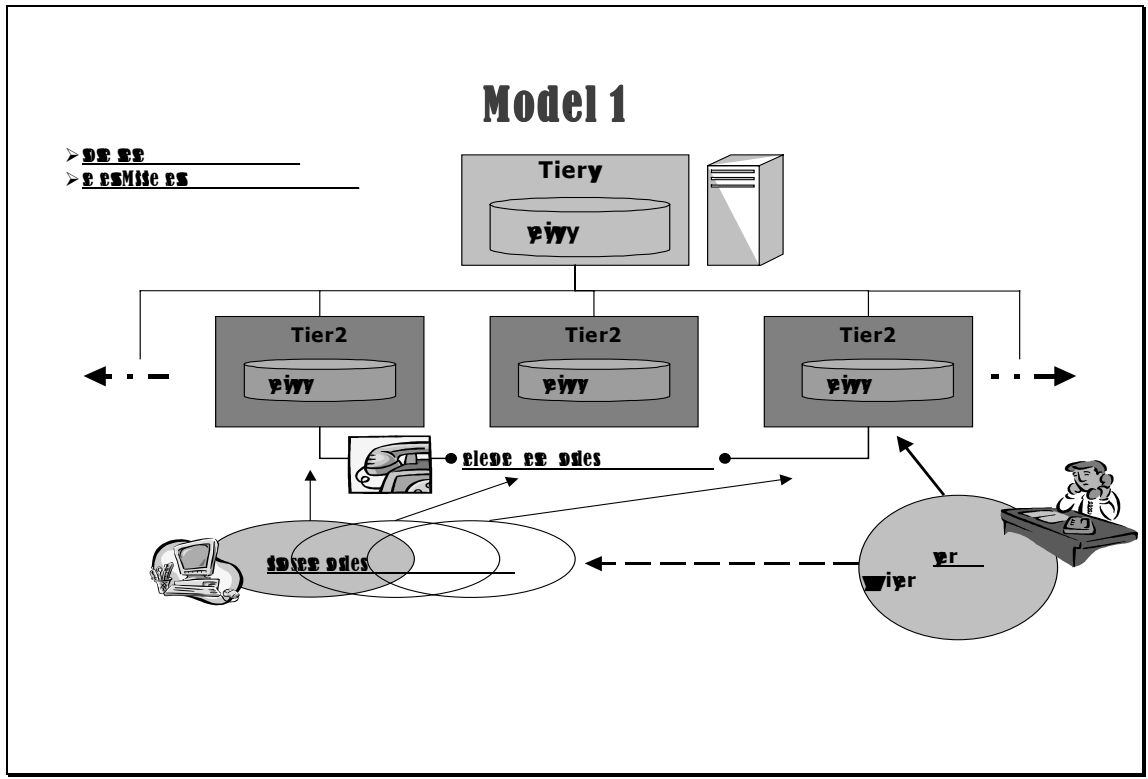

The interface between registry and registrar is used when the new registrar is provisioning its Resource Records (RRs) at the registry for telephone number. Also the registry can inform user change from old registrar to new one.

Protecting user's privacy might be well maintained among TSP registrars as long as there is no vertical integration of Tier 1 and 2 functions. The single registry service provider might implement policy on the access of private ENUM records to approve registrars.

The interface between registrar and the Application Service Provider (ASP) is used by ASP to provision, access and change the NAPTR RRs associated with the application services for the telephone number at the tier 2 if the user gives it the authorization to do so. When the user subscribes to any specific service or terminates it, the ASP informs the TSP. Users can subscribe to special ENUM services through ENUM ASPs, who will act as registrants to the TSPs. ASPs might unbundle the ENUM registrar service and let the user directly deal with the TSP's registrar service.

The interface between the new registrar and old registrar is used for resolving disputes about user's registrar change. The new registrar should inform the old registrar immediately when the new subscriber changes it, and also retrieve the NAPTR RRs at the old registrar.

The interface between registrar and user is used when the user requests or terminates to ENUM service from TSP, the registrar. The user also can authorize his/her ASPs for the access right to their NAPTR RRs at the registrar for a user's assigned telephone number.

The interface between ASP and user is used by ASP to provide its NAPTR RRs or the application service related information to user or to request the user to authorize to provision, access and change its NAPTR RRs at the registrar. Also, the user can subscribe or terminate application service(s) form ASP. Therefore, general ENUM service value flow of the model 1 is from users and ASP through TSP to Registry. The TSPs, and ASPs will have incentives operate ENUM provisioning services in this model. If the local telephony market among TSPs in the same region is competitive enough, registrar market can operate in a reasonably competitive manner.

### 4.2.2 Model 2

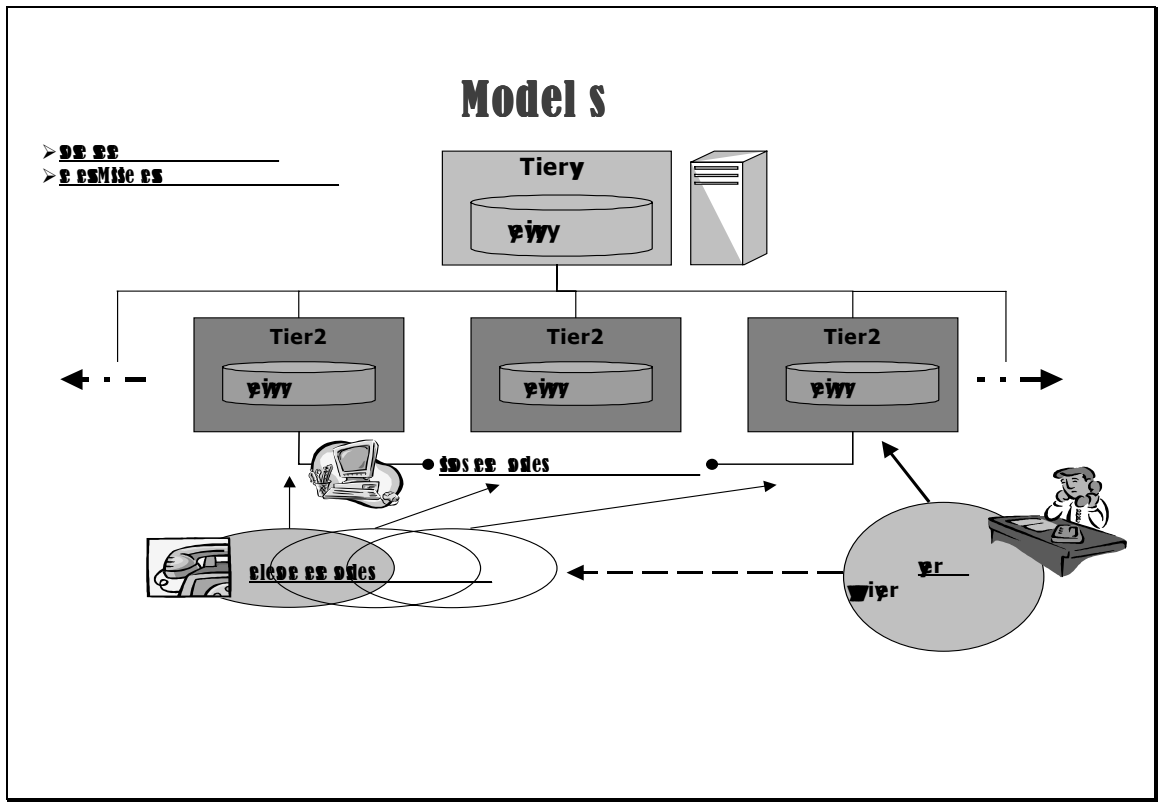

In this model, the Application Service Provider (ASP) plays a role as the registrar, and there could be other registrars that function equally. Most interfaces are same as *model 1*. The user must subscribe to TSP for telephony service and have the assigned telephone number. Those user who have telephone number assigned will get ENUM registrar service from ENUM ASPs (who provide ENUM enabled services) and those ASPs will pay the Registry for the ENUM service provisioning. For transparent operation of ENUM service in this model, there is an issue associated with the operation and cooperation from TSPs. There might be three approaches to motivate the operation of TSPs for ENUM provisioning domains; Registry directed, ASP directed and User directed. It might be TSP's business decision in this ENUM business model.

The interface between ASP and user of this model is same as interface of model 1 except the fact that it can have authorization directly from the user when he/she subscribes to ENUM service.

The interface between TSP and user of this model may be used to open a new telephony service account or to close it from the assigned telephone number.

When the user wants to subscribe to ENUM service, he/she may authorize the selected ASP to get all information about him/her from TSP. Also TSP can access the NAPTR RRs associated with the application services for the assigned user's telephone number if the user gives it the authorization to do so.

In this model, the role of TSP is somewhat smaller than that of model 1, though the TSP has its own principal role to assign a telephone number to the user for telephony service. From the user's view, he/she can select any ASP for ENUM service and that plays role as the registrar here without contacting with TSP if he/she already has the assigned telephone service. The ASP registrars should have appropriate authentication and access control mechanisms either through outsourcing or development to interact with the Tier 1 to ensure the user privacy protection.

General ENUM service value flow of this model is from users through ASPs to Registry and there might be independent economics and value flow from Registry, ASPs and users to TSPs depending on the business model of TSPs for ENUM services. This additional value stream will not only motivate the TSPs to cooperate with ASPs for ENUM provisioning but also promote the competition among TSPs in ENUM service provisioning domain.

### 4.2.3 Model 3

In this model 3, the independent registrar resides in tier 2 as the registrar. The independent registrar means a designated entity that plays role as a registrar and has relations with TSP, ASP and user for ENUM service. User might subscribe telephone service, ENUM enabled application and ENUM registration independently. However, the registry will collect the ENUM provisioning charging only from the independent Registrar. TSP and ASPs will establish the business relationship with Registrar if there is economic benefit of serving the end user.

For all interface in this model, each entity's principal function is same or very similar as those of model 1 and model 2. However, the interfaces among entities are more complex than other models, especially for users.

The interface between independent registrar and TSP/ASP may be used by TSP and ASP to provision, access and change the NAPTR RRs associated with the network related service and application service for the telephone number. Both of them can be done if the user gives TSP and ASP the authorization to do so.

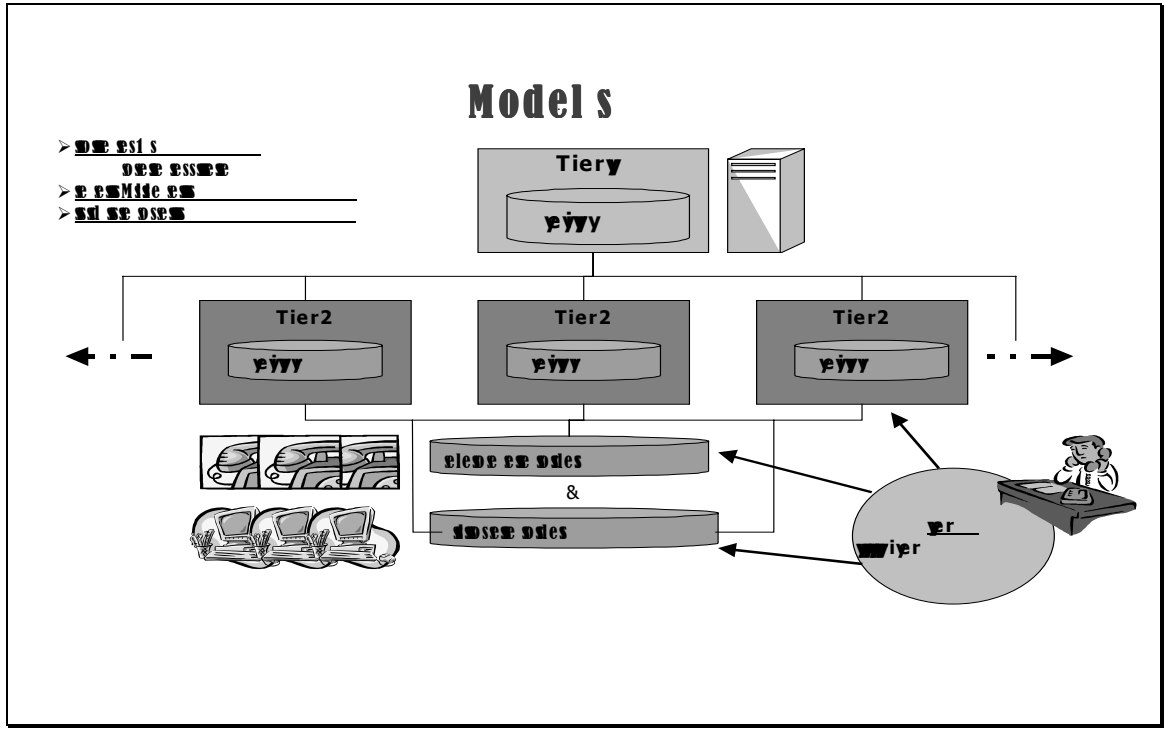

For the user privacy protection, it is essential that the single registry provider examine the approved independent registrars if they manage the user's privacy records correctly with cooperation with user's ASPs and TSPs. There will be emerging demand for the function to ensure that tier 2 registrars are properly handling the user privacy information with security.

In this model, the direction of ENUM provisioning service value flow is from user through registrar to the registry. ENUM ASP will have close business relationship with registrars since their main business is ENUM-based. TSP would have incentives to keep business relationship with the independent registrar, however probably not with all registrars.

### 4.2.4  Model 4

The different point of this model from previous models is that there are multiple registries. Again, the ENUM registrars of this model are the TSPs that have capability to manage the e164 number resource. It could be reasonable to allow multiple registries to promote competition for ENUM services. With multiple registries and registrars, the market formulation for ENUM services could be faster than that of one registry. The successful introduction of multiple ENUM registries will minimize any potential administrative bottlenecks and economic benefit can be achieved for the sake of the end users. The value flow among ENUM provisioning players will be same as in the Model 1 since the registry service should be transparent to the end users.

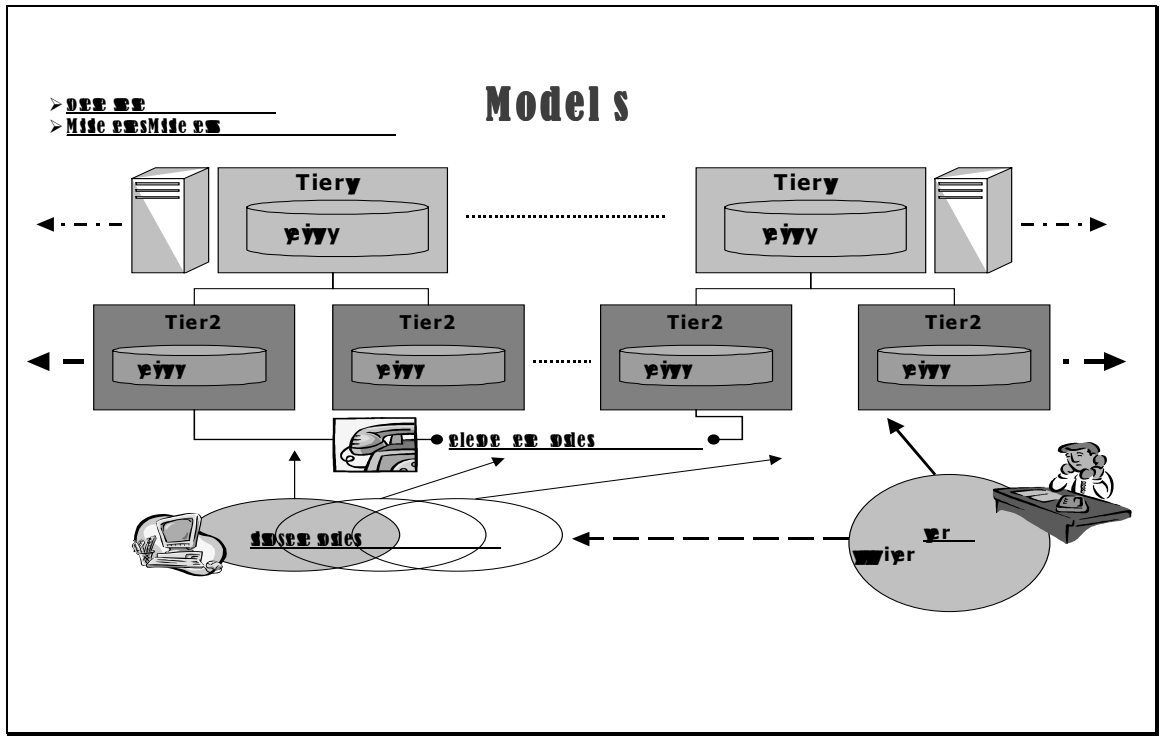

In this model, the relation and interface between registry and registrar are same as models of one registry. However, there should be the mutual connection among registries. For instance, if user A wants to communicate with user B who subscribe to ENUM service through TSP that is connected to different registry from user A's service, there should be clear interaction between different registries. When the user terminates ENUM service or change from old registrar to new one, the designated registry will inform the changed facts to other registries. The interface between TSP/ASP and user is same as in the model 1.

Privacy concerns may lead to vertical integration of registry and registrar functions among TSP players. Considering mergers and acquisition in the recent telephony market, this might be one of the most probable scenarios for TSPs in the ENUM registry and registrar service market with this multiple registry model.
Since TSPs will be in charge of registrar service, most of the ENUM customers of a TSP will be either its own telephone customers or customers who have interest in using e164 number that the TSP manages. Some phone-to-PC calls can be completed by using local e164 numbers through appropriate ENUM registration. This motivates both ASPs and TSPs to build ENUM enabled services with economic incentives. Different ENUM registries should be able to deal with pointer functions to different e164 provisioning ENUM numbers assigned to different TSPs. The question here is how we motivate these competing multiple registries to cooperate each other to share their information to seamless ENUM service. Are there any approaches to motivate the registries without segmenting their registrars into TSP's numbering areas? Complete registry service competition with cooperative resource provisioning may not be easy to achieve when the registrar service is provided by TSPs in this model. One potential solution for this problem would be creating additional value stream from either ASPs or ENUM service users to those registries. However, no such approach has been examined yet.

### 4.2.5 Model 5

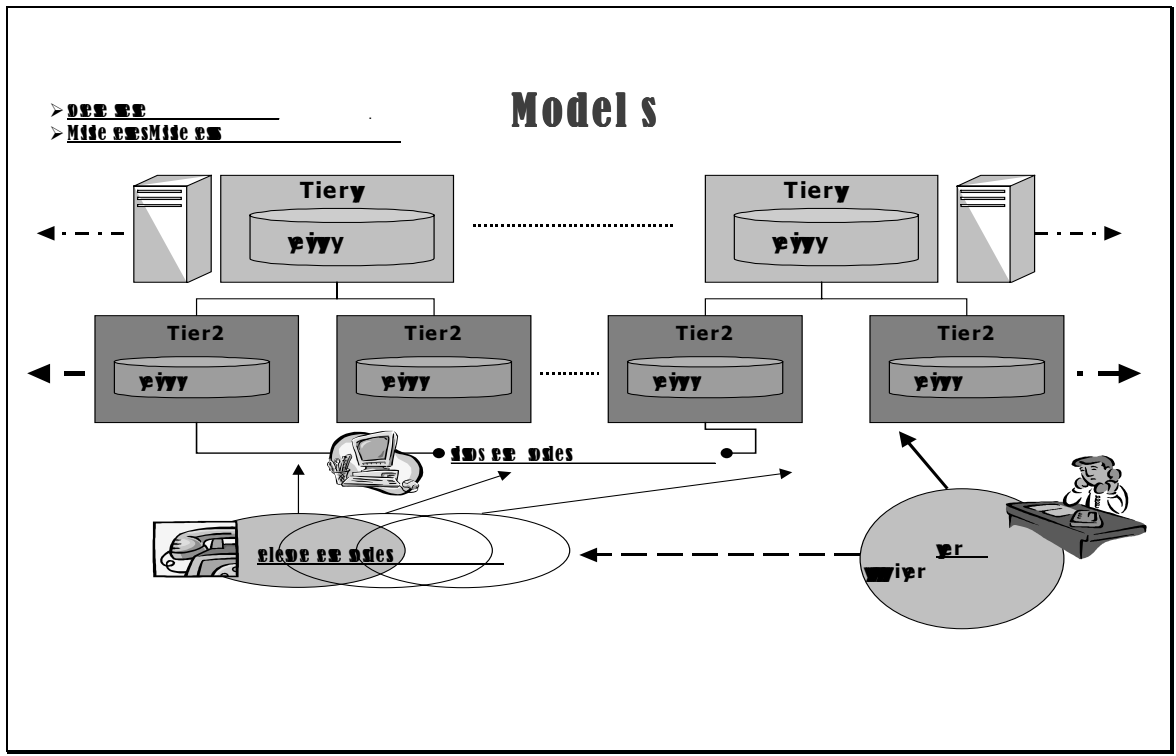

Like the case of *model 2*, the ASP plays a role as the registrar. The user will subscribe to TSP for telephony service and have the assigned telephone number. Those user who have telephone number assigned will get ENUM registrar service from ENUM ASPs (who provide ENUM enabled services) and those ASPs will pay the Registry for the ENUM service provisioning. The interfaces between ASP (or TSP) and user of this model will be similar to the one of model2.

In this model, the role of TSP and its interface to ASP registrar can be integrated with the TSP's relationship with the registries. For example, if the Tier 1 registry market is segmented based on the e164 numbering plan areas, the relationship between ASP Tier 2 and Tier1 registry will be arranged based on the user's e164 number subscription to the TSP. Still ASP registrar market can be competitive with ENUM enabled services even with such prearranged provisioning between Tier 1 and Tier2. This implies that all the ASP registrars need to keep their provisioning relationship with both all the Registries and TSPs of their customer. The other scenario can be the introduction of competitive registry market independent from the TSPs and let the ASP registrars to motivate the TSPs to cooperate for ENUM provisioning.

Different registries could form separate communities (possibly vertically integrated) with ASP registrars and TSPs to protect user's privacy ENUM information and handle it properly.  Also, it is required to have privacy provisioning between different registry communities for ENUM registry service in this model.

The ASP registrar's choice of registry will be it business and economic decision to acquire more customers with better ENUM enabled service.  In this competitive registry model, the ASP registrars would have an incentive to acquire e164 number management capability from the existing TSPs.

### 4.2.6 Model 6

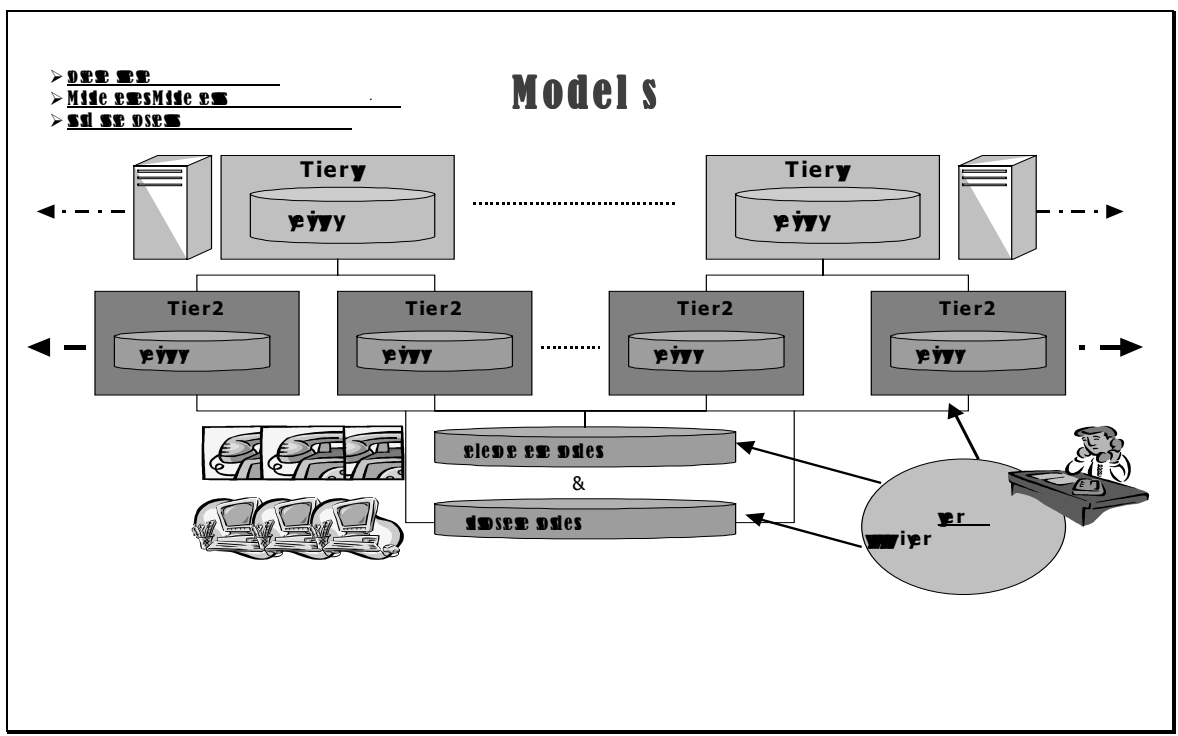

In this model 6, the independent registrar resides in Tier 2.  The Tier 2 registrars will compete among each other in the domain of price and feature registrar services.  Tier 2 independent registrars approach for choosing a registry for their service will be based on the service reliability and costs.  Many scenarios of Model 6 will be similar to model 5 except that Tier 2 may not need to pursue for the e164 management capability.   The independent registrars will eventually promote the competition among TSPs and ASPs for ENUM enabled service market.

The user privacy handling approach of this model would be similar to the approach of model 3 except that privacy provisioning requirements between among different registries.   There is little chance of forming rigid separate registry and registrar communities due to the separated market and lack of economic incentives to do so.

In this model, the TSPs and ASPs have incentives to cooperate with Tier 2 registrars and their registry for their ENUM enabled service markets. The registry service will be financed by charging the tier 2 independent registrars. The major question of this model is that whether there is enough market potential for multiple registries through only ENUM provisioning registry service. This question might be related to our original question whether we can number users and service instead of the telephone lines. More potential number possible, more likely the multiple-registry model will be successful.

**4.3 Implications of the Models**

In this section, we will analyze the different ENUM registrar player's strategy on ENUM administration model assuming the IP telephony is the major potential application for ENUM service.

First, TSPs (both ILECs and CLECs) would be supportive to the models of 1 and 4 where they have registrar presence on ENUM provisioning service domain. Probably, having single registry will secure their position to manage e164 numbering resources and the ENUM service provisioning. These models will allow TSPs to better realize the potential of the IP telephony market. TSPs would be able to create demand for many enhanced services and broadband access services that TSPs currently provide, and include new services for existing customers. According to Telecom ACT 1996, there would not be any firm restriction for TSPs to provide the ENUM registrar service as long as they are in competitive market in the telephone number domain at least for now. The supporting argument for TSPs for this position would be to ensure the integrity, reliability, and stability not only of the developing ENUM infrastructure, but also the existing global telephone numbering system. TSPs would like to become a Tier 2 provider to help provide ENUM registration services to interface and update customer records internally as their IP telephony service customer profile rather than dealing with other companies to provide the services. However, as other firms try to compete with the TSP in the ENUM registrar market, there will be pressures for open access and provisioning of the e164 number resource. In the long term, that will be major threat to TSPs.

Second, ASPs will have incentives to support the Model 5 where multiple registries provide motivations to compete with the cost advantages and enhanced services, including intelligent query responses to the ENUM database and service-handoffs can be bundled with ENUM registrar service. As we mentioned before, the ASPs will try to acquire the e164 resource management capability eventually. That will make situations where ASPs will have difficulties to get the cooperation from the TSPs for existing users with their numbers. Property of phone number needs to be examined in the context of ENUM registrar policy carefully, and it might be different from the "1-800" number model for "ownership" of a number. IP telephony service from these ASP registrars will complicate the issues of PSTN-IP telephony interconnection and other telecommunications policy.

Finally, independent registrars will be more supportive for the model 6 where Tier2 can find the economic registry service provider based on their economic decision.

Potentially, multiple registry model would promote more incentives for TSPs and ASPs to cooperate with registrars as long as the TSPs and ASPs are not directly competing with independent registrar in the registrar service market.  Vertical integration issue between ENUM registrar service and ENUM enabled application service (by either TSPs and ASPs) needs to be examined carefully to avoid any potential market failure in the ENUM provisioning service market.  The market opportunity and competition will be enhanced if there is a safeguard that ensures ENUM registrar service without vertical integration in the market of IP telephony.

## 5   Discussion and Conclusion

With the ENUM protocol still undergoing evaluation, we presented that there will be many opportunities in the IP telephony market.  This is largely because ENUM can be used with many of the other telecommunications services.  In order to gain a considerable customer base, the industry needs to find ways to lower the costs and promote the service development for the customers.  One approach can be bundling the services, however there are issues related to the vertical integration between ENUM provisioning service and ENUM enabled application services like IP telephony.   ENUM services have network externalities in the IP telephony market.  As the numbers of users increase in the future, additional services can be bundled to make ENUM an interoperable technology.

We consider service opportunities and its IP telephony market potential when ENUM service is created and successfully implemented. Without examining exact market and industry analysis, it may become more difficult and complex to set up a proper ENUM administration service model.  There might be a great deal of conflicts of interest among different ENUM players for new potential market generated by ENUM technology and services.  We examined the situations when TSPs, ASPs and independent registrars have incentives to provide ENUM registrar services for different market situations.

Regarding ENUM as service through the converged networking service, there are various emerging telecommunications policy issues and problems such as authentication, security, privacy, monopolistic infrastructure and so forth.

How will ENUM subscriber's privacy be protected?  Since ENUM will be designed using the architecture of the global Domain Name Services (DNS), it must be acknowledged that current implementation of DNS allows for open access of information to any party who requests it.  For the Internet to work as smoothly as it does an open DNS system is required, but this is quite the contrary with ENUM service.  ENUM handles more private information than DNS information.  To protect subscriber's personal information it is important that only their service providers have access to it, in order to implement such a mechanism ENUM might only contain pointers to a service context within a tier 2 provider's data store.  This will allow the subscriber's information to be protected and governed by laws and agencies within their service providers geographic region. This brings about another issue though, what rules and policies should govern personal information where there is no set policies, the answer to this question lies within the ENUM industry and public interests.

Privacy concerns may lead to vertical integration of the registrar and registry (Tier 1 and 2) functions. A service provider who wanted to guarantee greater security might choose not to run a registry open to all registrars, but limit access to its own or to specially approved registrars.

When attempting to guarantee consumer protection we must address the issues of authentication and validation. It is imperative that when allowing an end-user to populate their ENUM information that we guarantee they are the subscriber and not someone who is trying to hijack a number. If service providers cannot guarantee a subscriber's ENUM integrity, ENUM will be a failure. Security must be guaranteed both within the system and in end to end connections. To do this we must establish policies for authenticating end-users.

The model for Tier 1 registry market should be considered carefully. As we discussed previously, the model (single, multiple and competitive multiple) of Tier 1 would have important effects to the operation of ENUM service market and provisioning operation. It might be important to diagnose the potential public interests related ENUM registry service.

It is related to the telephone numbering services that the FCC might have the question of regulation or unregulation of the PSTN to DNS based ENUM service. The issue of concern to establish regulation might be closely related to the measures of public interests. Will the Commission regulate the ENUM registrars and registries and decide who the providers will be? Will it decide to allow TSPs to provide the ENUM service to consumers? Presently, the FCC has no legal basis to undertake any action in regards to regulating the ENUM service and IP telephony. [Cannon, 2001] [Frieden, 1997] However, FCC needs to watch for possible monopolistic behaviors once the market takes off.

The Telecommunications Act of 1996 presents a legal basis for individuals to enter into the business of communications. It also allows for competition in any market against others and it's intent is to reduce regulation. Sec 253 – Removal of Barriers to Entry – states that no restriction should be placed on entities (companies, individuals) trying to enter the market. This section of the Act provides for the unregulation of the tier level entries. Therefore, regulating the ENUM will be premature and may not be feasible right now. FCC's role as an ENUM representative member to the ITU-T presents an advantage as guidelines and policies continue to be developed. However, it is questionable whether we need a coordination body like ICANN for ENUM resources if there is enough interest from the public. This suggests that the regulatory issues raised by ENUM will not go away, and that regulators will have to continue to confront them as the market evolves.

In summary, we consider the following as our major interests in new ENUM services:

- The potential of ENUM to create new IP telephony service market,
- Administrative models of ENUM-based service (i.e., market organization) are analyzed based on the different players' incentives and motivations for market potential,
- The key policy implications of ENUM services are analyzed from the standpoint of the administrative model analysis, and
- Other regulatory policy issues are examined in the general context of ENUM issues.